# Polaron formation as origin of unconventional isotope effects in cuprate superconductors

A. Bussmann-Holder and H. Keller[2]
Max-Planck-Institute for Solid State Research, Heisenbergstr. 1, D-70569 Stuttgart, Germany
[2]Physik-Institut der Universität Zürich, Winterthurerstr. 190, CH-8057 Zürich, Switzerland

Various unconventional isotope effects have been reported in high-temperature superconducting copper oxides which are beyond the scheme of BCS theory. Their origin is investigated within polaron theory which leads to a renormalization of the single particle energies and introduces a level shift here. It is found that the exponential squeezing of the second nearest neighbour hopping integral carries the correct isotope effect on the superconducting transition temperature $T_c$, as well as the one on the penetration depth. The average superconducting gap is predicted to have an isotope effect comparable to the one on the penetration depth. The results imply that the coupling of the electronic degrees of freedom to the Jahn-Teller $Q_2$-type mode is the origin of these isotope effects.

High-temperature cuprate superconductors (HTSC) are one of most intensely studied systems due to the yet lacking understanding of the pairing mechanism. The antiferromagnetic properties of the undoped compounds are a consequence of the large Coulomb repulsion at the copper site. The energy scale given by it is the largest, and this has been taken as evidence that it must play a crucial role for the pairing mechanism. Consequently effects stemming from the lattice have mostly been ignored, especially in view of the fact that the isotope effect on the superconducting transition temperature $T_c$ almost vanishes at optimum doping [1,2]. The failure of BCS theory to account for many of the observed exotic properties has contributed to interpret the pairing mechanism in terms of a purely electronically driven one. However, various unexpected isotope effects have been reported [3,4,5] which are neither expected within the BCS mechanism nor within models based on strong correlations only. Since the Cu ion is one of the strongest Jahn-Teller systems [6], polaron formation can take place here and be the origin of unconventional isotope effects.

In spite of that the majority thought that the physics of high-temperature cuprate superconductors are dominated by a purely electronic mechanism, we address here the possibility of unconventional charge lattice coupling leading to Jahn-Teller polaron formation. We are motivated by various experimental findings as e.g. the observation of an isotope effect on the London penetration depth $\lambda_L$[3,4], the isotope dependence of the electronic energy bands [5], the interpretation of EPR data in terms of three spin polarons [7], the EXAFS data revealing the coexistence of different length scales[8], the isotope effect on $T_c$ [1,2], which exceeds the BCS value in the underdoped regime, the strain induced enhancement of $T_c$ in HTSC films [9,10]. In addition, we address the *inherent inhomogeneity* observed in HTSC as revealed by EXAFS [8], STM [11], NMR [12], EPR [13], which requires that a multi-component scenario has to be considered. Especially the observation of not only a d-wave order parameter but also an s-wave order parameter [14] is included in the modelling of HTSC.

For the undoped parent compounds we start with the t-J scenario where double occupation at the Cu site is forbidden because of the large Hubbard U [15]. With doping, this picture changes rapidly since <u>all</u> energy scales are destabilized and a substantial charge mismatch sets in which needs to be compensated for by *local polaronic* lattice distortions. Here also, antiferromagnetism is rapidly suppressed since the hole spin at the oxygen ion lattice site aligns antiparallel to the Cu ion spin [16] carrying the distortion with it. Thus a coexistence of locally distorted areas with the regular lattice is observed where the distortions are first randomly distributed, but in order to compensate for the large strain fields associated with them, organize into regular patterns as evidenced by "stripe" formation [17]. The ordering has the advantage that the antiferromagnetic matrix is partially preserved and that antiferromagnetic fluctuations are still "alive". In order to describe the physics of this system correctly, the antiferromagnetic background is subject to the physics of the t-J model, whereas the "extra" charge induces strong electron lattice coupling. Since the local lattice distortions around the extra charge are felt by the antiferromagnetic background, also here coupling to the lattice has to be incorporated. Consequently the t-J Hamiltonian is *extended* to incorporate the hole induced charge channel and the important effects from the lattice. This results in a two component Hamiltonian, where interactions between the charge channel (local hole plus induced lattice distortion) and the spin channel (antiferromagnetic fluctuations modified by lattice distortions) are explicitly included [18]:

$$H = H_{t-J} + H_{ch} + H_{ch-L} + H_{sp-L}$$

$$H_{t-J} = \sum_{i,s} E_{sp,i} d^+_{sp,i,s} d_{sp,i,s} + \sum_{i,j} T_{sp} n_{sp,i,\uparrow} n_{sp,j,\downarrow}$$

$$H_{ch} = \sum_{j,s} E_{ch,j} c^+_{ch,j,s} c_{ch,j,s} \quad [1]$$

$$H_{ch-L} = \sum_{k,q} [\mathbf{g}(q) b_q c^+_{ch,k+q} c_{ch,k} + h.c.]$$

$$H_{sp-L} = \sum_{k,q} [\mathbf{g}(q) b_q d^+_{sp,k+q} d_{sp,k} + h.c.]$$

Here the antiferromagnetic properties are taken into account in the t-J Hamiltonian $H_{t-J}$. In equ.1 $c^+$, $c$ and $d^+$, $d$ refer to creation and annihilation operators in the charge and spin p-d hybridised channels with $d^+d=n_{sp}$, $c^+c=n_{ch}$ and indices *ch* (charge), *sp* (spin); $E$ are site i, j dependent band energies with momentum k dependent dispersion

$$E_k = -2t_1(\cos k_x a + \cos k_y b) + 4t_2 \cos k_x a \cos k_y b - \mathbf{m},$$

where $a$, $b$ are the in-plane lattice constants, $t_1$, $t_2$ are nearest and second nearest neighbour hopping integrals, and $\mathbf{m}$ is the chemical potential which controls the number of particles and is the doping controlling parameter [19]. $T$ is the transfer integral between neighbouring spin up and down states avoiding double occupancy. $b_q$ are momentum q dependent phonon annihilation operators, and $\mathbf{g}$ is the charge/spin – lattice interaction constant which – for simplicity – is the same in both channels. Both bands do not cover the full Brillouin zone (BZ) but the charge band is confined to the antinodal directions over a small range of 9°, whereas the spin appears in the nodal directions with 40° weight in the first quadrant [20]. The important terms in the Hamiltonian are those proportional to the electron – phonon coupling constant $\mathbf{g}$. These are well known with manifold consequences. They can induce: i) a charge density wave instability [21] accompanied by a structural phase transition which is not observed in HTSC; ii) BCS type superconductivity, which – however – shows a doping independent isotope effect and no isotope effect on $\lambda_L$; iii) a Peierls transition with spin or charge ordering [22], where again both would be accompanied by a structural instability; iv) a combined charge – spin – density wave [23]; which excludes the appearance of superconductivity; v) polaron formation [24] which neither inhibits superconductivity nor is in contrast to the observed isotope effects. We thus concentrate on the last possibility and apply standard techniques to diagonalize equs. 1 [24]. This leads to renormalizations of all involved energy scales [18], but the most relevant ones occur in the electronic band energies:

$$E_{sp,ch} = -2t_1 \exp[-\mathbf{g}^2 \coth \frac{\hbar \mathbf{w}}{2kt}](\cos k_x a + \cos k_y b)$$
$$+ 4t_2 \exp[-\mathbf{g}^2 \coth \frac{\hbar \mathbf{w}}{2kt}]\cos k_x a \cos k_y b - \Delta^* - \mathbf{m} \quad [2]$$

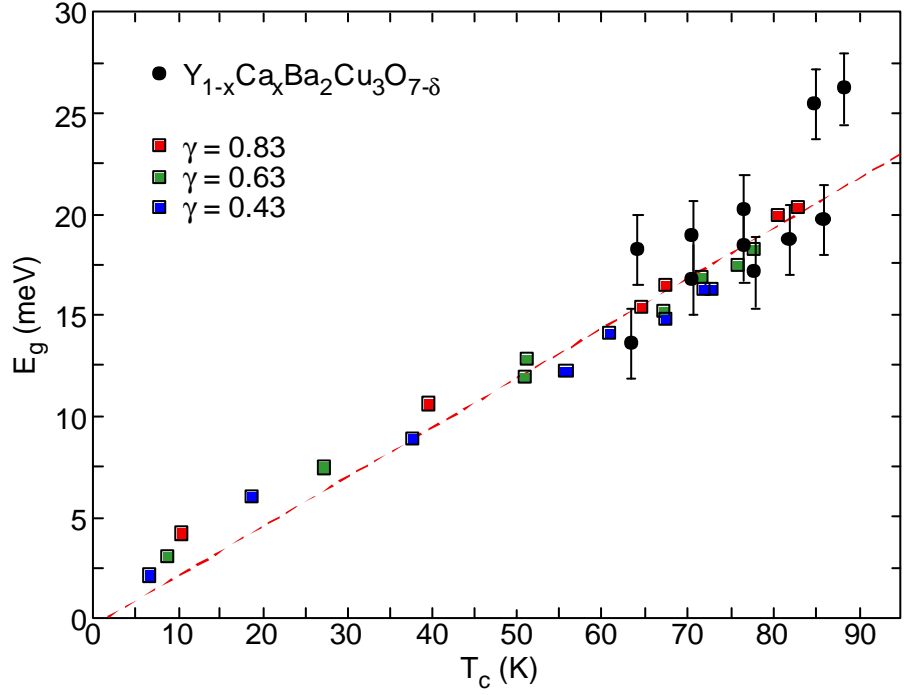

**Figure 1**: The average superconducting energy gap $E_g = \sqrt{E_{g,d}^2 + E_{g,s}^2}$ as a function of the superconducting transition temperature $T_c$. Squares are calculated values with $\gamma$=0.43, 0.63, 083 (blue, green, red), respectively, whereas black symbols are experimental data points for $Y_{1-x}Ca_xBa_2Cu_3O_{7-\delta}$ taken from Ref. 14.

Here $\omega$ is a characteristic phonon frequency and $\Delta^*$ the polaronic level shift, proportional to the local lattice displacements. Most importantly a squeezing (band narrowing) effect on both hopping integrals appears which is proportional to $g^2$ and is inversely dependent on the square root of the ionic mass. Superconducting properties of the coupled charge – spin – lattice system are studied within an effective two-band Bogoliubov quasiparticle approach, where pair wise attractive interactions in the spin channel induce d-wave superconductivity, whereas the interaction in the charge channel is too weak to cause superconductivity. However, due to an attractive phonon mediated inter-channel interaction superconductivity is induced there as well [18]. The resulting scenario is a two gap superconducting state, analogous to $MgB_2$ [25], with the distinction that the order parameters are of different symmetries. In addition, and opposite to $MgB_2$, in cuprates mostly a time-averaged gap is observed caused by fast fluctuations. The superconducting gaps $E_{g,s}$ (s-wave gap), $E_{g,d}$ (d-wave gap) have been calculated as a function of doping for various coupling constants $\gamma$ and fixed ratio of $t_2/t_1$=0.3 as suggested by band structure calculations for YBCO [26]. Figure 1 shows the average gap $E_g = \sqrt{E_{g,s}^2 + E_{g,d}^2}$ as a function of the corresponding $T_c$ where experimental data points for $Y_{1-x}Ca_xBa_2Cu_3O_{7-\delta}$ [14] have been added for direct comparison.

Furthermore, we have calculated the oxygen-isotope ($^{16}O/^{18}O$) effects on $E_g$ and $T_c$ for various values of $\gamma$ and the same parameters as in Fig 1. As shown in Fig. 2, the relative isotope shifts $|\Delta E_g/E_g| = |(^{18}E_g - ^{16}E_g)/ ^{16}E_g|$ and $|\Delta T_c/T_c| = |(^{18}T_c - ^{16}T_c)/ ^{16}T_c|$ are found to be equal (red dashed line). This striking finding is in excellent agreement with experimental data [27] of the

oxygen-isotope effect on the the zero-temperature in-plane magnetic penetration depth $\lambda_{ab}(0)$ and $T_c$ also in included in Fig. 2 for comparison.

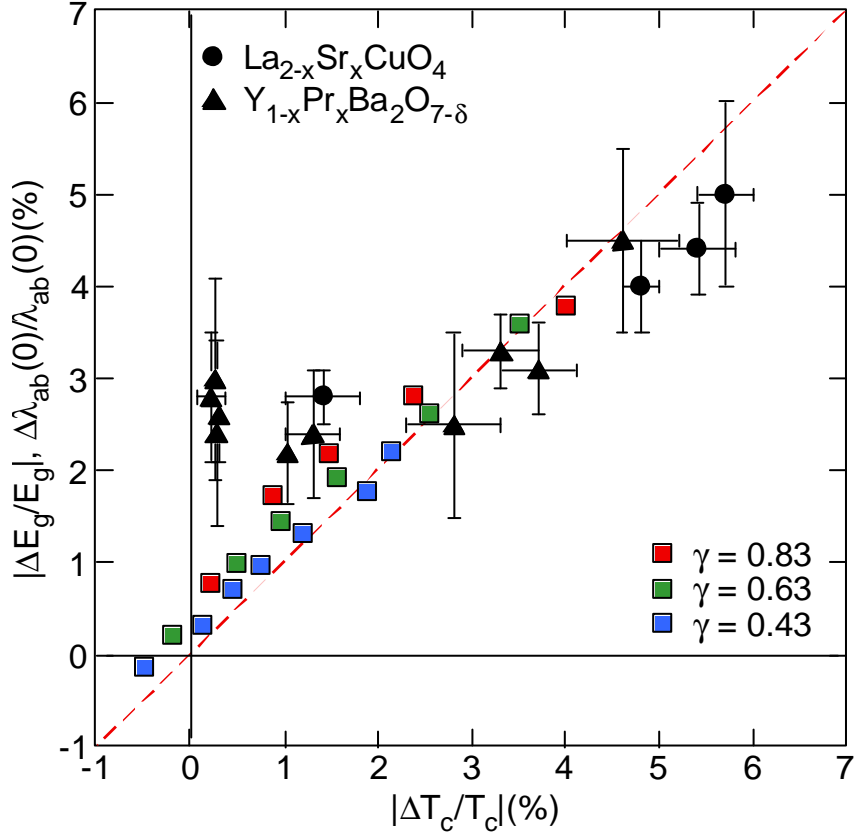

**Figure 2**: Relative isotope shift of the average gap $E_g = \sqrt{E_{g,d}^2 + E_{g,s}^2}$ ($|\Delta E_g/E_g|$) as a function of the relative isotope shift of the superconducting transition temperature $T_c$ ($|\Delta T_c/T_c|$). Both gaps, the s-wave gap $E_{g,s}$ and the d-wave gap $E_{g,d}$ show nearly the same isotope shift. Squares are calculated values with $\gamma$=0.43, 0.63, 083 (blue, green, red), respectively. The dashed red line is a guide to the eye. Black symbols refer to experimental oxygen-isotope effect data of the zero-temperature in-plane magnetic penetration depth $\lambda_{ab}(0)$ and $T_c$ taken from Ref. 27.

However, the saturation effect observed in $\Delta\lambda_{ab}(0)/\lambda_{ab}(0)$ at optimum doping is not found for $|\Delta E_g/E_g|$ and appears only if $t_1/t_2$ is strongly enhanced as compared to the present value. The close resemblance between $|\Delta E_g/E_g|$ and $\Delta\lambda_{ab}(0)/\lambda_{ab}(0)$ in Fig. 2 is not accidental since in both quantities the leading term for the isotope effect stems from the band energies. Interestingly, a similar linear relation has recently been reported for the band energy isotope effect as a function of the gap values [5]. Further we have investigated the variation of the oxygen-isotope effect on $T_c$ with doping in comparison to experimental data. This is shown in Fig. 3 where the calculated isotope effect exponent $a = -\partial \ln T_c / \partial \ln m$ is shown as a function of $T_c/T_{c,max}$, together with experimental data [27] for comparison ($T_{c,max}$ is the maximum $T_c$ for a particular

family of HTSC). With decreasing doping ($T_c/T_{c,max}$) $a$ systematically increases from nearly zero at optimum doping to the BCS value of $a\approx0.5$ in the underdoped regime. In the overdoped regime $\alpha$ changes sign and adopts negative values. This prediction is hoped to be verified experimentally soon. But, it is supported by the isotope effect on the band energies [5] which also changes sign.

In order to clarify the symmetry of the coupling lattice distortion which causes these isotope effects, the following analysis has been performed: i) first only $t_1$ is renormalized by the polaronic coupling whereas $t_2$ is bare; ii) only $t_2$ is renormalized and $t_1$ remains unrenormalized. The results are included in Fig. 3. Clearly, the isotope effect due to $t_1$ only, deviates strongly from experimental observations in the underdoped regime where it approaches zero. On the other hand $t_2$ follows the total isotope effect correctly. From this finding we conclude directly about the lattice distortion which governs the isotope dependence of the gaps, $T_c$ and $\lambda_{ab}(0)$. The half-breathing mode (Fig. 4, left panel), which shows anomalous softening[28], is dominated by $t_1$, and obviously carries the wrong isotope dependence.

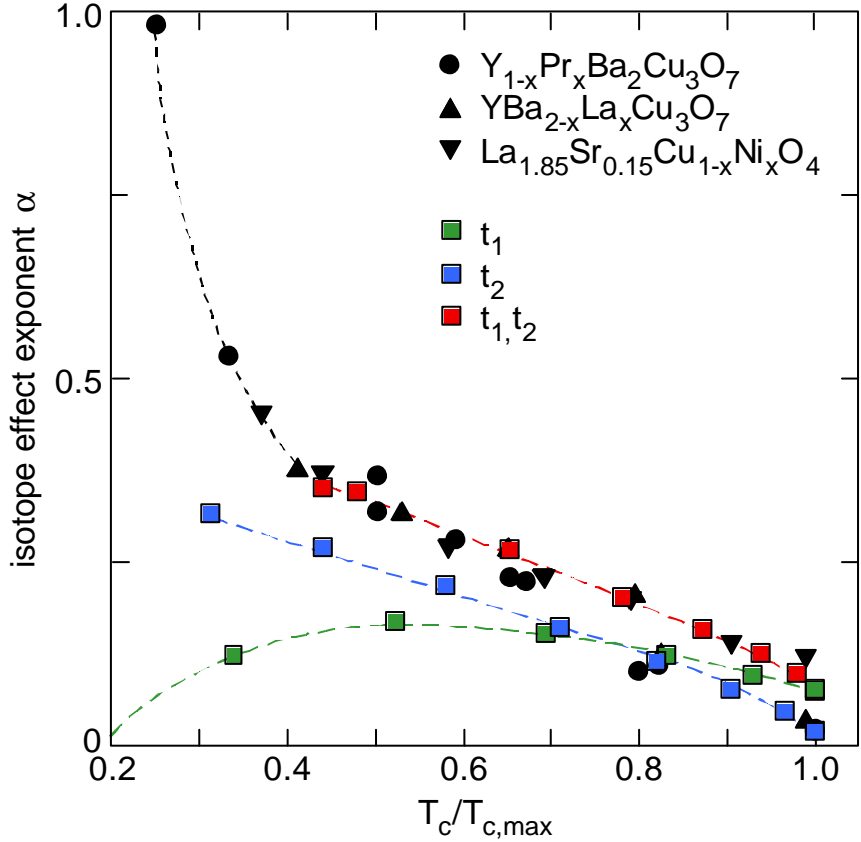

**Figure 3** Calculated isotope effect exponent $\alpha$ as a function of $T_c/T_{c,max}$ for $\gamma=0.83$ ($T_{c,max}$ is the maximum $T_c$ for a given family of HTSC). The red squares are calculated by renormalizing both hopping elements $t_1$, $t_2$ through the polaronic coupling proportional to $\gamma^2$ (equ. 2). The green squares are calculated by renormalizing $t_1$ only, whereas for the blue squares $t_2$ is renormalized, $t_1$ remains bare. The black symbols are experimental data points for various HTSC taken from Ref. 27. The dashed lines are a guide to the eye.

Since symmetry considerations also apply to the perpendicular direction of the half breathing mode, also the full breathing mode can be excluded. The crucial role of $t_2$ can only be taken into account by considering the Jahn-Teller active $Q_2$ type mode as the origin of the observed effects (Fig. 4, right panel). Our explanation is consistent with the interpretation of EPR data [7], but also with data for perovskite type manganites[29], where the dominant role of the Jahn-Teller formation has been demonstrated by isotope experiments.

Since the discovery of HTSC [30] was motivated by the idea that Jahn-Teller polaron formation could be a new electron (hole) pairing mechanism, the above results get back to the discovery and support the original ideas. Also similar considerations should be taken into account for cobaltites where currently rather exotic pairing mechanisms are discussed.

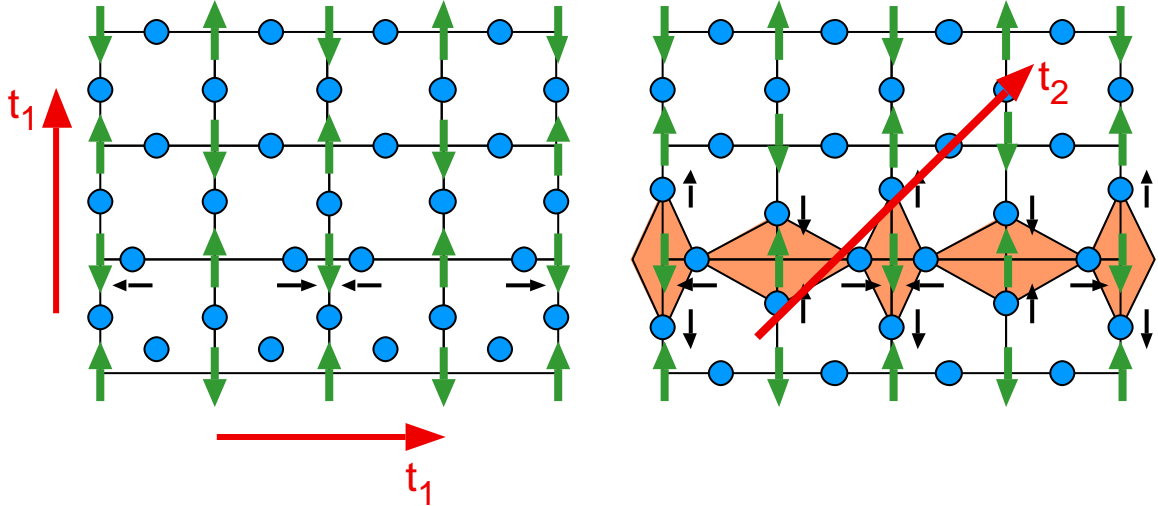

**Figure 4**: The relevant ionic displacements which are either governed by $t_1$ only (left panel) or by $t_2$ only (right panel). Here only displacements in the $CuO_2$ plane are considered. The blue circles represent the oxygen ions, the green arrows are copper ions with the antiferromagnetic order. The black arrows in the left panel indicate the displacements for the LO half breathing phonon mode. Similarly a full breathing mode could be governed by $t_1$. The coloured rhombohedra show the displacements (black arrows) of the $Q_2$ type mode which is dominated by $t_2$.

**References**

1. J. P. Franck, in *Physical Properties of high temperature superconductors IV* (ed. Ginsberg, D. M.) 189-293 (World Scientific, Singapore, 1994)
2. D. Zech, et al. *Nature* **385**, 681-683 (1994)
3. G.-M. Zhao, M. B. Hunt, H. Keller & K. A. Müller, *Nature* **385**, 236-239 (1997)
4. R. Khasanov, et al. *Phys. Rev. Lett.* **92,** 057602 1-4 (2004)
5. G. H. Gweon, et al., *Nature* **430**, 187-189 (2004)
6. K. A. Müller, in *Magnetic resonance and relaxation* (ed. Blinc, R.) 192-208 (North Holland Pbl. Inc., 1966)
7. B. I. Kochelaev, J. Sichelschmidt, B. Elschner, W. Lemor & A. Loidl, *Phys. Rev. Lett.* **79**, 4274-4277 ((1997)
8. A. Bianconi et al., *Phys. Rev. Lett.* **76**, 3412-3415 (1996)
9. J.-P. Locquet, et al., *Nature* **394**, 453-456 (1998)
10. M. Abrecht, et al., *Phys. Rev. Lett.* **91**, 057002 1-4 (2003)
11. S. H. Pan, et al., *Nature* **413**, 282-285 (2001)
12. J. Haase & C. P. Slichter, *J. Supercond.* **16**, 473-475 (2003)
13. A. Shengelaya, et al., *Phys. Rev. Lett.* **93**, 017001 1-4 (2004)

14. A. Kohen, G. Leibowitch, & G. Deutscher, *Phys. Rev. Lett.* **90**, 207005 1-4 (2003)
15. P. W. Anderson, *Science* **235**, 1196-1198 (1987)
16. F. C. Zhang & T. M. Rice, *Phys. Rev. B* **37**, 3759-3761 (1988)
17. J. M. Tranquada, B. J. Sternlieb, J. D. Axe, Y. Nakamura & S. Uchida, *Nature* **375**, 561-563 (1995)
18. A. Bussmann-Holder, et al., *J. Phys.: Cond. Mat.* **13**, L168-L171 (2001)
19. K. M. Shen, et al., *Cond-mat/*0407002
20. Bianconi, A. Private comm.
21. T. M. Rice & G. K. Scott, *Phys. Rev. Lett.* **35**, 120-123 (1975)
22. E. Pytte, *Phys. Rev. B* **10**, 4637-4642 (1974)
23. S. N. Bahara & S. G. Mishra, *Phys. Rev. B* **31**, 2773-2775 (1985)
24. S. G. Lang & Yu. A. Firsov, *Sov. Phys. JETP* **16**, 1302-1312 (1963)
25. H. J. Choi, D. Roundy, H. Sun, M. L. Cohen & S. G. Louie, *Nature* **418**, 758-760 (2002)
26. E. Pavarini, I. Dasgupta, T. Saha-Dasgupta, O. Jepsen & O. K. Andersen, *Phys. Rev. Lett.* **87**, 047003 (2001)
27. R. Khasanov, et al., *J. Phys.: Cond. Mat.,* in press (2004), and refs. therein
28. M. Tachiki, M. Machida & T. Egami, *Phys. Rev. B,* **67**, 174506 1-12 (2003)
29. G.-M. Zhao, K. Condor, H. Keller & K. A. Müller, *Nature* **381**, 676-678 (1996)
30. J. G. Bednorz & K. A. Müller *Z. Phys.: Cond. Mat.* **64**, 189-193 (1986)

**Acknowledgements**: It is a great pleasure to acknowledge helpful discussions with K. A. Müller, A. R. Bishop, A. Simon, D. Pavuna and R. Micnas. This work was partly supported by the Swiss National Science Foundation and by MaNEP.